\documentstyle[preprint,aps,psfig]{revtex}
\begin{document}
\tightenlines
\draft
\title{Dynamical restriction for a growing neck due to mass
parameters in a dinuclear system}
\author{G.G.Adamian$^{1,2,3}$, N.V.Antonenko$^{1,2}$, A. Diaz-Torres$^{1}$
and W.Scheid$^{1}$}
\address{$^{1}$Institut f\"ur Theoretische Physik der
Justus--Liebig--Universit\"at,
D--35392 Giessen, Germany\\
$^{2}$Joint Institute for Nuclear Research, 141980 Dubna, Russia\\
$^{3}$Institute of Nuclear Physics, 702132 Tashkent, Uzbekistan}
\date{\today}
\maketitle

\begin{abstract}
Mass parameters for collective variables of a dinuclear system
and strongly deformed mononucleus are microscopically formulated with
the linear response theory making use of
the width of single particle states and the fluctuation--dissipation
theorem. For the relative motion of the nuclei
and for the degree of freedom describing the neck between the nuclei,
we calculate mass parameters with basis states of the
adiabatic and diabatic two-center shell model.
Microscopical mass parameters are found
larger than the ones obtained with the hydrodynamical model and
give a strong hindrance for a melting of the dinuclear system along
the internuclear distance into a compound system.
Therefore, the dinuclear system lives a
long time enough comparable to the
reaction time for fusion by nucleon transfer.
Consequences of this effect for the complete fusion process are discussed.

\end{abstract}


\section{Introduction}
The synthesis of transuranium and superheavy elements,
the production of superdeformed nuclei and  nuclei far from the
line of stability stimulate the study of
phenomena of complete fusion in
heavy ion collisions at energies less than 15 MeV/nucleon.
The fusion of heavy ions is characterized by
the formation of a dinuclear system (DNS) at the initial stage
of reaction when a significant part of the
kinetic energy is transferred into excitation energy.
In that case one has to describe the evolution
of the DNS and strongly deformed
nucleus by collective coordinates $Q$, namely by
the distance $R$ between the centers of colliding nuclei
(or relative elongation $\lambda$ of the system), the mass and charge
asymmetry degrees of freedom $\eta=(A_1-A_2)/A$
and $\eta_Z=(Z_1-Z_2)/Z$, respectively, ($A_1$,
$A_2$ and $Z_1$, $Z_2$ are mass and charge numbers of the nuclei,
$A=A_1+A_2$, $Z=Z_1+Z_2$), the parameter $\varepsilon$ of the
neck and deformation coordinates of nuclei
\cite{1,2,3,4,5,V1,6,7,8}.
Then dissipative, conservative and inertial
forces for collective variables, arising from
the nucleus-nucleus interaction, are needed to be determined.

The present fusion models can be
distinguished by the choice of the relevant collective
variables along which fusion mainly occurs. While many models
study the fusion in $R$ at a
practically fixed value of $\eta$, the DNS model ~\cite{6,9} considers
the DNS evolution in mass asymmetry by nucleon or cluster transfers
as the main path to the
compound nucleus. The DNS model assumes basically that the neck degree of
freedom is fixed in the evolution in $\eta$ and
the nuclei are hindered to melt together
by a variation in the relative distance.
As was shown in Refs.~\cite{10,11,MN,12}
this occurs due to the relatively large inertia of the neck degree of
freedom and structural forbiddenness effects.
In this paper we study dynamical restrictions for the growth of the
neck in the DNS and suggest proper methods of
the calculation of the DNS inertia tensor.

There are various  macroscopical and microscopical approaches to calculate
the inertia tensor \cite{B1}.
The macroscopical approaches
(see, for example, \cite{5,8}) are based on the hydrodynamical model of the
nucleus.
A calculation of the inertia tensor with a theory for quantum fluid dynamics
is suggested in Ref.~\cite{13}. By using a random-matrix
model to describe the coupling between a collective
nuclear variable and intrinsic degrees of freedom and with the help of the
functional integral approach, mass parameters
are derived in Ref.~\cite{W1}.
In the linear response theory \cite{H1,Iva}
the inertia tensor is found for fissioning nuclei.
The microscopical approaches mainly use the cranking type expression
and perform calculations in  different single particle bases applying
adiabatic \cite{3,4,S1} or diabatic \cite{15,16} two-center shell models.
Difficulties in the cranking type calculations arise for
collective motions with large amplitudes, for example, in fusion or fission,
due to pseudo--crossings or crossings of levels
in the single particle spectrum.
Some publications disregard the contributions
from the crossings (pseudo--crossings)
which means a neglect of
effects of configuration changes
on the mass parameters during the evolution of the nuclear shape
in spite of the fact that the collective inertia
is strongly influenced by level crossings
(pseudo--crossings) \cite{P1,G1}.
In order to overcome this problem, two--body
collisions should be regarded which
lead to a width of the single particle levels and an
effective reduction of the level crossing effects.
For example, calculations of the nuclear inertia
in a generalized cranking model with pairing correlations
yielded masses of about one order
of magnitude larger than the ones without pairing \cite{T1}.

One of the aims of the present paper is to obtain analytical
expressions for mass parameters using models and methods
which include residual interaction effects.
In Sect.~2 the mass parameters are
obtained within the linear response theory taking the
fluctuation-dissipation theorem and
the width of single particle states
into account. The same mass parameters are also derived
by Fermi's golden rule and by smoothing the single particle spectrum in
the mean--field cranking formula.
In Sect.~3 the mass parameters
for the relevant collective variables (mass asymmetry, elongation,
neck and deformation parameters) of the
DNS and strongly deformed nuclear systems are
evaluated in the two-center shell model with adiabatic and diabatic bases.
Sect.~4 contains a short summary.

\section{Microscopical inertia}
\subsection{Derivation from collective response function}
Let us consider a nuclear system described by a single collective
coordinate $Q$ and intrinsic single particle coordinates $x_i$
(with the conjugated momentum $p_i$) and assume the following
effective Hamiltonian \cite{H1}:
\begin{eqnarray}
\hat H(x_i,p_i,Q)&=&\hat H(x_i,p_i,Q_0) + (Q-Q_0)\hat F(x_i,p_i,Q_0)\nonumber\\
&+&\frac{1}{2}(Q-Q_0)^2<\frac{\partial ^2 \hat H(x_i,p_i,Q)}
{\partial Q^2}>_{Q_0,T_0}
\end{eqnarray}
The shape of the nuclear mean field is changed with the
collective coordinate $Q$
that introduces the coupling between $Q$ and the nucleonic
degrees of freedom.
Eq.~(1) is obtained by expanding the Hamiltonian to second order
in the vicinity of $Q_0$ and by approximating the second order
term in a kind  of unperturbed limit.
The coupling term between the collective
and intrinsic motion is proportional to the first order in
$\delta Q=Q-Q_0$
with an operator $\hat F$ given
by the derivative of the mean field with respect to $Q$ in the neighborhood
of $Q_0$.
In this way global motion is described within a locally harmonic
approximation.

By applying the linear response theory \cite{H1,K1},
the Fourier transform of the collective response function is
defined as
\begin{eqnarray}
\chi_{coll}(\omega)=
\frac{\chi(\omega)}{1+k\chi(\omega)},
\end{eqnarray}
where
$\chi(\omega)$ is the Fourier transform of the response
function for intrinsic motion and
measures how, at some given $Q_0$ and temperature $T_0$,
the nucleonic degrees of freedom react to the coupling term
$\hat F\delta Q$.
The coupling constant $k$ is written in the form
\begin{eqnarray}
-k^{-1}=
<\frac{\partial^2 \hat H(x_i,p_i,Q)}
{\partial Q^2}>_{Q_0,T_0}=
\frac{\partial^2 E(Q,S_0)}
{\partial Q^2}|_{Q_0} + \chi(\omega=0)=
C(0)+\chi(0)
\end{eqnarray}
with $\chi(0)$ and $C(0)$ being the static response
and stiffness, respectively.
Since the constant $k$ is entirely determined by quasi-static properties,
it is no surprise that $E$ is
the internal energy at a given entropy $S_0$ or the free energy
at a given temperature $T_0$.
The structure of Eq.(2) reflects self--consistency
between the treatment of collective and microscopic dynamics.
It expresses the response of the system of interacting nucleons
in terms of the response of the individual nucleons.

The local motion in the $Q$ variable is described in terms of the
collective response function $\chi_{coll}$.
A short calculation leads to
the following approximation for the mass coefficient \cite{H1,Iva,K1,H3}:
\begin{eqnarray}
M= - \frac{1}{2k^2}\frac{\partial ^2 (\chi_{coll}(\omega))^{-1}}
{\partial \omega^2}|_{\omega=0}=
(1+\frac{C(0)}{\chi(0)})^2[M^{cr}+\frac{\gamma^2(0)}{\chi(0)}],
\end{eqnarray}
where
\begin{eqnarray}
M^{cr}= \frac{1}{2}\frac{\partial ^2 \chi (\omega)}
{\partial \omega^2}|_{\omega=0}
\end{eqnarray}
is the inertia in the zero-frequency limit of the second
derivative of the intrinsic response
function. $M^{cr}$ can be shown to be similar to the
one of the cranking model. For many applications,
the value of $C(0)/\chi(0)$ is much less than unity. The additional term
$\gamma^2(0)/\chi(0)$ in Eq.~(4) gives
a positive contribution to $M$ where $\gamma (0)$ is
the friction coefficient defined by
\begin{eqnarray}
\gamma (0)=
-i\frac{\partial  \chi (\omega)}{\partial \omega}|_{\omega=0}
=\frac{\partial  \chi^{''} (\omega)}{\partial \omega}|_{\omega=0}
=\frac{1}{2T_0}\psi^{''}(0).
\end{eqnarray}

The dissipative part of the response function $\chi^{''}(\omega)$
is connected with the
dissipative part of the correlation function $\psi^{''}(\omega)$
through the fluctuation-dissipation theorem:
\begin{eqnarray}
\chi^{''}(\omega)=\frac{1}{\hbar}
\tanh (\frac{\hbar\omega}{2T_0})\psi^{''}(\omega).
\end{eqnarray}
The $\psi^{''}(\omega)$ has a singularity of $\delta$--function type at
$\omega=0$:
\begin{eqnarray}
\psi^{''}(\omega)=2\pi\psi^{0}\delta(\omega) + \psi^{''}_R(\omega)
\end{eqnarray}
with  $\psi^{''}_R(\omega)$  being regular at $\omega=0$.
In the case of an independent particle model we have
\begin{eqnarray}
\psi^{''}(\omega)=\pi\hbar\sum_{j,k}^{}|F_{jk}|^2n(e_j)[1-n(e_k)]
[\delta(\hbar\omega-e_{kj}) + \delta(\hbar\omega+e_{kj})].
\end{eqnarray}
Here, $e_{kj}=e_k-e_j$ is the difference of single particle energies,
$n(e_{j})$ are the occupation numbers
and $F_{jk}=<j|\hat F|k>$ the single particle matrix elements
of the operator $\hat F$.
At $j=k$  and $\omega=0$ we find
the contributions from the diagonal matrix elements:
\begin{eqnarray}
\psi^{0}=\sum_{k}^{}|F_{kk}|^2n(e_k)[1-n(e_k)]
= T_0\sum_{k}^{}\left|\frac{\partial n(e)}{\partial e}\right|_{e=e_k}
\left(\frac{\partial e_k}{\partial Q}\right)^2.
\label{ksi0_eq}
\end{eqnarray}
The last part in (\ref{ksi0_eq}) was derived with a
Fermi distribution for the occupation numbers, which is characterized by the
temperature $T_0$.
The value of $T_0$ does not effectively go to zero with decreasing excitation
energy because each single particle level has a width
due to the two-body interaction. Indeed at zero excitation
energy the distribution of the occupation numbers deviates
from a step function at least due to pairing correlations.
If we replace the $\delta$--functions in Eq.(8) or (9),
we have to apply a Lorentzian functions with the double
single particle width $2\Gamma$ where $\Gamma$ is the single particle
width, because $\hbar\omega$ is the transition energy between
two single particle states \cite{H1}.
Therefore, we substitute the $\delta-$functions in Eq.~(9) by
$\Gamma/[\pi((\hbar\omega\pm e_{kj})^2+\Gamma^2)]$. Then using
Eqs.~(6)-(10), we can write the friction coefficient in the following
form:
\begin{eqnarray}
\gamma(0)=\gamma^{diag}(0) + \gamma^{nondiag}(0),
\end{eqnarray}
where
\begin{eqnarray}
\gamma^{diag}(0)=
\frac{\hbar}{\Gamma}\sum_{k}^{}
\left|\frac{\partial n(e)}{\partial e}\right|_{e=e_k}
\left(\frac{\partial e_k}{\partial Q}\right)^2.
\end{eqnarray}
For smaller temperatures $T_0 < 2$ MeV which are of
interest here, $\gamma^{diag}(0)$ is much larger than $\gamma^{nondiag}(0)$
\cite{H1}.
The static response is found as
\begin{eqnarray}
\chi(0)=
\lim_{\epsilon\to 0}\int\limits_{-\infty}^{+\infty}
\frac{d\omega}{\pi}\frac{\chi^{''}(\omega)}{\omega-i\epsilon}
=\lim_{\epsilon\to 0}\int\limits_{-\infty}^{+\infty}
\frac{d\omega}{\hbar\pi}\frac{\tanh(\frac{\hbar\omega}{2T_0})
\psi^{''}(\omega)}{\omega-i\epsilon}=
\chi^{diag}(0)+ \chi^{nondiag}(0),
\end{eqnarray}
where
\begin{eqnarray}
\chi^{diag}(0)=
\sum_{k}^{}\left|\frac{\partial n(e)}{\partial e}\right|_{e=e_k}
\left(\frac{\partial e_k}{\partial Q}\right)^2.
\end{eqnarray}
With realistic assumptions $\gamma^{diag}(0) \gg\gamma^{nondiag}(0)$ and
$\chi^{diag}(0)\gg\chi^{nondiag}(0)$ and neglecting
$C(0)/\chi(0)$, we can divide the mass parameter (4) as
\begin{eqnarray}
M=M^{diag} + M^{nondiag}.
\end{eqnarray}
The contribution
of the diagonal matrix elements of $\hat F$ to $M$ are
\begin{eqnarray}
M^{diag}=\frac{(\gamma^{diag}(0))^2}{\chi^{diag}(0)}
=\frac{\hbar^2}{\Gamma^2}\sum_{k}^{}
\left|\frac{\partial n(e)}{\partial e}\right|_{e=e_k}
\left(\frac{\partial e_k}{\partial Q}\right)^2.
\label{MD_eq}
\end{eqnarray}
If the single particle widths are properly taken into account,
the nondiagonal contributions to the inertia are \cite{S1}
\begin{eqnarray}
M^{nondiag}=M^{cr}
=\hbar^2\sum_{k\ne k'}^{}\frac{|F_{kk'}|^2}{e^2_{kk'}+\Gamma^2}
\frac{n(e_k)-n(e_{k'})}{e_{k'}-e_k}.
\label{MND_eq}
\end{eqnarray}
The main contribution to $M$ is the diagonal part $M^{diag}$
because it dominates for collective variables which are responsible
for changes of the nuclear shape of the system \cite{P1,G1,GR1}.
Note that the calculation of $M^{diag}$ is simpler than
$M^{nondiag}$.
For the case that the pairing residual interaction is regarded
and only  diagonal matrix elements
in the cranking formula are taken into account, Eq.~(\ref{MD_eq})
was obtained with $\Gamma=\Delta$ ($\Delta$ is the pairing gap)
in Ref.~\cite{1,T1}. Starting with an equation for
the single particle density matrix extended with an approximate
incorporation of particle collisions in the relaxation
time approach, the authors of Ref.~\cite{Ivan}
derived an expression similar to (\ref{MD_eq})
(with $\Gamma=\hbar/\tau$, $\tau$ is the
relaxation time) but with a negative sign. This negative sign
arises from the fact that the condition of self-consistency
between collective and nucleonic
dynamics was disregarded in ~\cite{Ivan}  which is important for a
correct calculation of the mass parameters.
It was stressed in \cite{H1,H4}  that
within the linear response theory the diagonal
component of the friction parameter originates from the "heat pole" of
the correlation function and vanishes when the system is ergodic.
As shown in Ref.~\cite{KITAY}, the well necked DNS-type configurations
are not ergodic and stable against chaos.
Even at zero excitation energy the level crossings at the Fermi surface
lead to considerable mass flow \cite{P1,G1,GR1} and the diagonal component
of the correlation function (or mass parameter) does not vanish.

Besides the mass  and friction coefficients, the
diffusion coefficients $D_{kl}$ $(k,l=(Q,P))$
must also have a component diagonal in the
matrix elements of $\hat F$ because they are connected
with correlation functions. For example, the diffusion coefficient in momentum
is defined as:
\begin{eqnarray}
D_{PP}=\frac{1}{2}\psi^{''}(\omega=0)=T_0\gamma(0).\nonumber
\end{eqnarray}

\subsection{ Derivation from Fermi's golden rule}
By setting  $\hat H_I=(Q-Q_0)\hat F(x_i,p_i,Q_0)$
as perturbation (see Eq.~(1)), the decay rate of a collective
state $|n>$ with energy $E_n$ to the collective state $|m>$ with
energy $E_m$ is given in lowest order according to
Fermi's golden rule:
\begin{eqnarray}
w(n\to m+\hbar\omega)
&=&\frac{2\pi}{\hbar}|<m|Q-Q_0|n>|^2\nonumber\\
&\times&\int\limits_{}^{}d(\hbar\omega)
|<\hbar\omega|\hat F|0>|^2
\delta (E_n-E_m-\hbar\omega)\rho_{qs}(\hbar\omega).
\label{WNM_eq}
\end{eqnarray}
Here, the integral is taken over the final states of the intrinsic system
with the density $\rho_{qs}$.
With the fluctuation-dissipation theorem for small temperatures
we have \cite{Iapona}:
\begin{eqnarray}
|<\hbar\omega|\hat F|0>|^2
\rho_{qs}(\hbar\omega)d(\hbar\omega)=
\frac{2}{\pi}\frac{\hbar\omega}{2}R(\omega)d\omega
\end{eqnarray}
with the relaxation function
$R(\omega)={\chi^{''}(\omega)}/{\omega}$.
The half-decay width is obtained from Eq.(\ref{WNM_eq}) as
\begin{eqnarray}
\Gamma_n=\hbar\sum_{m}^{}w(n\to m+\hbar\omega).
\label{GAM_eq}
\end{eqnarray}
Using the properties of the response function and a Taylor expansion
\begin{eqnarray}
R(\omega)=\frac{\chi^{''}(\omega)}{\omega}=
\frac{1}{\omega}\left[\chi^{''}(\omega=0) +
\left.\frac{\partial \chi^{''}(\omega)}{\partial \omega}
\right|_{\omega=0}\,\omega
+ \left.\frac{1}{2}\frac{\partial^2
\chi^{''}(\omega)}{\partial \omega^2}
\right|_{\omega=0}\,\omega^2+...\right],
\end{eqnarray}
and the standard formula for mass $M$ ($E_n\ne E_m$)
\begin{eqnarray}
M=\frac{\hbar^2}{2}(\sum_{m}^{}|<m|Q-Q_0|n>|^2[E_n-E_m])^{-1}
\label{Summr_eq}
\end{eqnarray}
we obtain by setting $\Gamma=\Gamma_n$ with
Eqs.~(\ref{WNM_eq})--(\ref{Summr_eq})
\begin{eqnarray}
M=\frac{\hbar}{\Gamma}
\frac{\partial \chi^{''}(\omega)}{\partial \omega}|_{\omega=0}
=\frac{\hbar}{\Gamma}\gamma(0).
\label{M22_eq}
\end{eqnarray}
Large temperatures in (19) effectively lead
to a temperature dependence of $\Gamma$ in (\ref{M22_eq}). Since
$\gamma(0)$ in Eq.~(11) contains
the terms with the diagonal matrix elements of
the operator $\hat F$,
the mass parameter $M$ also has the
diagonal component $M^{diag}$ (\ref{MD_eq}).
So, the contributions to the mass parameter
can be again classified as those with diagonal and nondiagonal
matrix elements, respectively.

That the mass parameter is proportional to the friction coefficient
(see Eq.~(\ref{M22_eq})), has an analogy in the hydrodynamic model.
For multipole moments $\lambda$ of the nucleus
with $\lambda > 1$, the following ratio in the limit of an
irrotational flow was derived in \cite{33}
$$ M^{irr}_{\lambda}/\gamma^{irr}_{\lambda}=
\frac{3A^{2/3}}{8\pi(2\lambda+1)(\lambda-1)r_0}\frac{1}{\nu},$$
where $\nu$ is the coefficient of the two-body viscosity
and $r_0=1.2$ fm.

\subsection{Derivation from the mean--field cranking formula}
Using the single particle spectrum and the corresponding wave functions,
one can obtain the mass parameter with the cranking formula
\begin{eqnarray}
M^{cr}= \hbar^2\sum^{}_{\alpha\ne \beta}
|<\alpha|\frac{\partial}{\partial Q}|\beta>|^2
\frac{n(e_{\alpha}) - n(e_{\beta})}{e_{\beta} - e_{\alpha}}.
\label{Cr1_eq}
\end{eqnarray}
In reality the Hamiltonian of the system
contains a residual two-body interaction
between the nucleons in addition to the mean field. The
residual coupling distributes the strength of single particle
states over more complicated states. This
spectral smoothing causes the effect that
the sum over $\alpha$ and $\beta$  appearing in
(\ref{Cr1_eq}) also includes diagonal terms with
$\alpha=\beta$. Let us prove this statement.

The Eq.(\ref{Cr1_eq}) can be rewritten as
\begin{eqnarray}
M^{cr}= \hbar^2\sum^{}_{\alpha\ne \beta}
\int\limits_{}^{}de_1\delta (e_1-e_{\alpha})
\int\limits_{}^{}de_2\delta (e_2-e_{\beta})
|<e_1|\frac{\partial}{\partial Q}|e_2>|^2
\frac{n(e_1) - n(e_2)}{e_2 - e_1}.
\label{Cr21_eq}
\end{eqnarray}
Next we use the following replacements
\begin{eqnarray}
\int\limits_{}^{}de_1g(e_1)\to\sum_{k_1}^{},\quad
\int\limits_{}^{}de_2g(e_2)\to\sum_{k_2}^{},
\end{eqnarray}
\begin{eqnarray}
\delta(e-e_k)\to\rho_{k}(e)=\frac{1}{2\pi}
\frac{\Gamma}{(e-e_k)^2+(\Gamma/2)^2}
\end{eqnarray}
and the approximation
\begin{eqnarray}
D^2\,|<k_1|\frac{\partial}{\partial Q}|k_2>|^2\approx
|F_{k_1k_2}|^2.
\end{eqnarray}
Here, $D=1/g$ is average energy distance between single particle states.
Then we express the mass (\ref{Cr21_eq}) as
\begin{eqnarray}
M^{cr}= \frac{\hbar^2\Gamma^2}{4\pi^2}\sum^{}_{\alpha\ne \beta,k_1,k_2}
\frac{|F_{k_1k_2}|^2
}{[e_{\alpha k_1}^2+(\Gamma/2)^2]
[e_{\beta k_2}^2+(\Gamma/2)^2]}
\frac{n(e_{k_1}) - n(e_{k_2})}{e_{k_2} - e_{k_1}}.
\label{Cr3_eq}
\end{eqnarray}
The energy spreading of the single particle states
is taken into account in Eq.(\ref{Cr3_eq}).
A line broadening happens if collisions of particles and holes with
the background result in single particle and single hole strength functions
that are concentrated around the original single particle energies.
The quantity $D\rho_{k}$ determines the average
strength function for a particle in state $k$ \cite{BM}.
In the limit $e_{k_1}=e_{k_2}=e_{k}$,
for $e_{\alpha,\beta}\approx e_k$, i.e. when two
neighboring levels near the level
$k$ are considered,
and $8/\pi^2\approx 1$, Eq.(\ref{Cr3_eq}) leads to Eq.(\ref{MD_eq}).
Thus, diagonal terms in the mass parameters appear because of
the finite width of the single particle levels due to
the residual interaction.

\section{Results of calculations}
\subsection{Adiabatic two-center  shell model}
Since in fusion and quasifission we deal with strongly
elongated systems, the two-center shell model (TCSM)
\cite{3,4,10} is appropriate
for calculating the potential energy surface.
In the TCSM the nuclear shapes are defined by the following
set of the coordinates: The elongation $\lambda=l/(2R_0)$ measuring
the length $l$ of the system in units of the diameter $2R_0$ of the
spherical compound nucleus and used to
describe the relative motion, the mass and charge asymmetry coordinates
$\eta$ and $\eta_Z$, respectively,
the neck parameter $\varepsilon=E_0/E'$  defined
by the ratio of the actual barrier height $E_0$ to the barrier
height $E'$ of the two-center oscillator, and
the deformation parameters $\beta_i=a_i/b_i$, $i=1,2$, of axially
symmetric fragments, defined
by the ratio of the semiaxes of the fragments.
Since collisions above the Coulomb barrier
are discussed, we firstly consider spherical nuclei with $\beta_i=1$
and then analyze the deformation effects.

In order to calculate the width of the single particle states,
we use the expression \cite{H1,J1}
\begin{eqnarray}
\Gamma_{k}=\frac{1}{\Gamma_0}
\frac{\left(e_{k} - e_{F}\right)^2+ (\pi T_0)^2}
{1+[\left(e_{k} - e_{F}\right)^2+ (\pi T_0)^2]/c^2}.
\label{CrG1_eq}
\end{eqnarray}
Here, $e_F$ is the Fermi energy. Both parameters $\Gamma_0$ and
$c$ are known from experience with the optical model potential
and the effective masses \cite{H1}.
Their values are in the following ranges:
$0.030\,$MeV$^{-1}\le {\Gamma_0}^{-1} \le 0.061\,$MeV$^{-1}$,
$15\,$MeV$\le c \le 30\,$MeV.
For small excitations, Eq.~(\ref{CrG1_eq}) is reduced
to the expression known in the theory of a Fermi liquid.
Since each single particle state has its own width,
 Eq.~(\ref{MD_eq})  is generalized as
($Q_i=\lambda,\eta,\eta_Z,\varepsilon,\beta_1,\beta_2$):
\begin{eqnarray}
M^{diag}_{ij}= \hbar^2\sum^{}_{k}
\frac{f_{k}}{\Gamma^2_k}
\frac{\partial e_{k}}{\partial Q_i}
\frac{\partial e_{k}}{\partial Q_j}.
\label{Cr2_eq}
\end{eqnarray}
For the Fermi occupation numbers $n(e_{k})$, the function
\begin{eqnarray}
f_{k}=
-\frac{dn_k}{de_k}=
\frac{1}{4T_0}
\cosh^{-2}\left(\frac{e_{k} - e_F}{2T_0}\right)
\end{eqnarray}
has a bell-like shape with a width $T_0$ and is peaked
at the Fermi energy $e_F$.

Various calculations of the mass parameter for the motion in $\lambda$
were carried out with expressions similar to Eq.~(\ref{Cr2_eq}),
for example in [1,20,25].
When the system adiabatically moves towards the
compound nucleus, the value of
$M_{\lambda\lambda}$ approximately increases by a factor 10--15 in
our and other calculations. In this paper we concentrate on the
calculation of the mass parameter $M_{\varepsilon\varepsilon}$ for
the motion of the neck to test whether the DNS exists sufficient time
enough with a relatively small neck.
The dependence of $M_{\varepsilon\varepsilon}$ on $\varepsilon$
is presented in Fig.~1 for the system $^{110}$Pd+$^{110}$Pd
at $\lambda=1.6$ which corresponds to the touching configuration
in this symmetric reaction.
The obtained values of $M_{\varepsilon\varepsilon}$
have the same order of magnitude as in [20]
where the pairing correlations were
taken into consideration.
The value of $M_{\varepsilon\varepsilon}$
increases by a factor 2.5 when the system falls into the fission--type
valley \cite{10}. This increase reflects the decrease of the shell
correction $\delta U$ with $\varepsilon$ towards $\varepsilon\to 0$.
Smaller values of $\delta U$ correspond to larger masses.

In order to obtain the same TCSM potential energy as in the DNS model
for the touching configuration,
the neck parameter $\varepsilon$ should be set about 0.75 \cite{10}.
With this value of $\varepsilon$
the neck radius and the distance between the centers of the
nuclei are approximately equal to the corresponding quantities
in the DNS.

For the parameter $c$ in Eq.~(\ref{CrG1_eq}) we use the "standard"
value 20 MeV since the masses (31) depend only
weakly on this parameter.
Setting the parameter
$\Gamma^{-1}_0=0.045\,$MeV$^{-1}$
in (\ref{Cr2_eq}) and comparing our results with
$M^{WW}_{ij}$ obtained in the Werner-Wheeler approximation, we find
$M_{\lambda\lambda}=M^{WW}_{\lambda\lambda}$,
$M_{\varepsilon\varepsilon}\approx (20-30)M^{WW}_{\varepsilon\varepsilon}$,
$M_{\lambda\varepsilon}\approx 0.4M^{WW}_{\lambda\varepsilon}$
and
$M_{\lambda\varepsilon}
/\sqrt{M_{\lambda\lambda}M_{\varepsilon\varepsilon}}\ll 1$,
practically independent of the mass number of the system.
Therefore, we can conclude that the
microscopical mass parameter of the neck
is much larger than the one in the Werner-Wheeler approximation and
the nondiagonal component $M_{\lambda\varepsilon}$ is small.
Since at the touching configuration
the slope of the single-particle levels is small and changes slowly with
decreasing elongation, the microscopical mass parameter in
$\lambda$ is close to its smooth, hydrodynamical value. In contrast,
a large amount of internal reorganization, which occurs at level crossings
with decreasing  $\varepsilon$,
leads to a large neck inertia of the initial DNS.
So, the value of $M_{\varepsilon\varepsilon}$ exceeds the mass
in the hydrodynamical model due to large values of
$|\partial e_k/\partial \varepsilon|$.
The restriction for the growth of the neck may be understood
by analysing the single particle spectrum as a function
of $\varepsilon$ ~\cite{10}.
For large $\varepsilon$ and
well necked-in shapes, the single particle spectrum shows
a good shell structure. The levels show an increasing number of
level crossings with increasing $\varepsilon$.

The time-dependence of the neck parameter calculated
with the microscopical and Werner-Wheeler mass formulas are
compared in the upper part of Fig.~2. The lower part of Fig.~2 shows
trajectories in the $(\varepsilon,\lambda)$-plane, calculated
with microscopic and Werner-Wheeler masses.
An adiabatic potential energy surface is used
in all these calculations.
Since there are no suitable barriers
at smaller values of $\lambda$ and $\varepsilon$
in the adiabatic potential which hinder a growth of the neck,
the neck parameter and system length decrease steadily to smaller values,
faster in the case with the Werner-Wheeler masses and
much slower with the microscopical masses.
The experiments on fusion of heavy nuclei can not be explained
as a melting with increasing neck together with a decreasing $\lambda$ in
an adiabatic potential \cite{10}.
It seems that there is an intermediate situation
between the adiabatic and diabatic
limits. The study of the transition between diabatic and adiabatic regimes
gives a potential energy surface which contains quite high barriers for the
motion to smaller $\lambda$ and $\varepsilon$ \cite{MN,M1}. Therefore,
the dynamical calculations with the adiabatic potential energy show a
maximal possible growth of the neck.
Due to the large moment of inertia in
the heavy nuclear system,
the dependence of potential energy on the angular momentum in dynamical
calculations is disregarded here. Indeed, in fusion reactions
with massive nuclei only low
angular momenta ($<$ 20 $\hbar$) contribute to the evaporation residue
cross section. For the study of fusion, the isotopic composition
of the nuclei forming the DNS is chosen with the condition of a
$N/Z$-equilibrium in the system with $\eta_Z$ following the value of
$\eta$ [10]. In processes developing in the shorter times,
the coordinate $\eta_Z$ could be
considered as independent collective variable.

As result of Fig.~2 we find that the
microscopical mass parameters keep the system
near the entrance configuration for a sufficient long time comparable with
the time of reaction even in an adiabatic potential.
Therefore, this situation justifies the assumption of a
fixed neck as we assume it in the DNS model \cite{6,9}.
When the DNS configuration exists a long time, then thermal
fluctuations in the mass asymmetry coordinate play the essential role
in the fusion process.
Indeed these fluctuations are responsible for the fusion
in the DNS model \cite{6,9}.
Thus, the dynamical restriction for the growth of the neck can be
caused partly by a large microscopical mass parameter for the neck motion,
partly by the potential energy surface intermediate between
diabatic and adiabatic limits.

In the adiabatic two-center shell model the mass parameter
$M_{\varepsilon\varepsilon}$ slightly increases with the
deformation parameters $\beta_i$ of the two nuclei in the symmetric
system $^{110}$Pd+$^{110}$Pd as shown in Fig.~3. The small variation of
$M_{\varepsilon\varepsilon}$ is due to the shell structure.
Therefore, the relatively large value of $M_{\varepsilon\varepsilon}$
is a general result for collisions of both spherical and deformed nuclei.
Fig.~4 shows the mass parameter $M_{\varepsilon\varepsilon}$
at the touching configuration of symmetric systems
$\eta=0$ with $\beta_i=1$ as a function of the mass number $A$.
The mass $M_{\varepsilon\varepsilon}$
increases with the mass of the system. It slightly decreases with increasing
mass asymmetry of the DNS.

\subsection{Diabatic two-center  shell model}
Let us now consider the calculation of the mass parameters (\ref{Cr2_eq}) with
the diabatic single particle energies obtained with the
method of maximum-symmetry in the diabatic two-center shell model \cite{16,M1}.
In the diabatic motion the nucleons do not occupy the lowest
single particle states as in the adiabatic case, but remain in their
diabatic states.

Due to the nonzero
widths of the single particle states, the distribution of
single particle strength over more complicated states is a Lorentzian
distribution $\rho_{k }(e)$  as introduced in Eq.~(27)
instead of the $\delta$-function $\delta (e-e_{k})$ \cite{H1}.
The occupation number $\widetilde{n}(e_{k })$ of a state $k$ with
energy $e_{k }$
and the corresponding value of  $\widetilde{f}_{k \text{ }}$ are
obtained, respectively, from functions $n(e)$ and $dn(e)/de$
calculated with zero widths of the levels
\begin{equation}
\widetilde{n}(e_{k})=\int n(e)\rho _{k}(e)de,
\end{equation}
\begin{equation}
\widetilde{f}_k=-\int \frac{dn(e)}{de}\rho _{k}(e)de.
\label{5}
\end{equation}
The Lorentzian distribution
increases the diffuseness of the Fermi-distribution.
The Fermi distribution which is given at
the touching configuration of the nuclei in the DNS is destroyed
if the further motion of the system runs diabaticly.
To treat the diabatic case, we use the following function $n(e)$
for an arbitrary configuration of the system
\begin{equation}
n(e)=\sum_{l=0}^{N}a_{l}\left( \Theta (e-e_{l})-\Theta (e-e_{l+1})\right),
\label{ne_eq}
\end{equation}
where $\Theta (x)$ is the Heavyside's function and
$e_{l}$ the energy of single particle state $l$ with
the occupation number $a_{l}$.
Here, the numbers $l=0,...,N$ count the single particle
states in the region of the Fermi level.
The values $e_{0}$ and $e_{N+1}$ are the low
and high limits of the single particle energies.
For lower and higher energies, the occupation
numbers are one and zero, respectively.
Therefore, we assume $a_{0}=1$ and $a_{N}=0$ in (\ref{ne_eq}).
The derivative $dn(e)/de$ is expressed as
\begin{equation}
-\frac{dn(e)}{de}=(1-a_{1})\delta
(e-e_{1})+\sum_{l=2}^{N-1}(a_{l-1}-a_{l})\delta (e-e_{l})+a_{N-1}\delta
(e-e_{N}).  \label{7}
\end{equation}

Then we obtain $\widetilde{f}_{k}$ with (\ref{5}) as follows
\begin{equation}
\widetilde{f}_{k}=(1-a_{1})\rho _{k
}(e_{1})+\sum_{l=2}^{N-1}(a_{l-1}-a_{l})\rho _{k }(e_{l})+a_{N-1}\rho
_{k}(e_{N}).
\label{8}
\end{equation}
In the calculations we assume the same average width
for each Lorentzian $\rho_{k}(e)$.
The diabatic occupation numbers $a_{l}$ are fixed at the touching
configuration of the system using a Fermi distribution for a
smaller temperature $T_0^{*}<T_0.$
The value of $T_0^{*}$ is chosen in such a way
that the occupation numbers $\widetilde{n}(e_{k})$
and the values of $\widetilde{f}_{k}$
obtained at the touching configuration of the nuclei with
the Fermi distribution or the
approximate expressions (\ref{ne_eq})-(\ref{8}) are nearly equal.

It is known that with zero width of single-particle levels
the mass parameters are smaller in the diabatic case than
in the adiabatic motion [18,22].
We found
values for the masses
$M_{\lambda\lambda}$  and $M_{\varepsilon\varepsilon}$ within the
diabatic two-center shell model approach which
are close to the ones obtained in adiabatic calculations
because we take the width
$\Gamma_k$  of single-particle states into consideration.
The finite decay widths $\Gamma_k$
destroy partially
the diabatic motion  (diabaticity).
As result, the diabatic distribution of
the single-particle occupation numbers approaches
the adiabatic Fermi-distribution for
the fixed values of $\varepsilon$ and $\lambda$
and excitation energy of system.
The mass parameters at the touching configuration are the same in the adiabatic
and diabatic descriptions  because the single-particle levels
and occupation numbers are practically the same in  these two limits.
As in the adiabatic two-center shell model,
the mass parameters $M_{\lambda\lambda}$ and $M_{\varepsilon\varepsilon}$
depend weakly on the mass asymmetry $\eta$ and on the deformations $\beta$
of the nuclei
($\beta=\beta_1=\beta_2$ for $^{110}$Pd+$^{110}$Pd)
at the touching configuration. This result is presented in Fig.~5.
Due to the regard of  the width of the single-particle states,
the mass parameters depend more smoothly on the mass asymmetry $\eta$
than the mean field cranking masses.
A stronger  dependence of the mass parameters on
$\eta$ is expected for  entrance channels
with a larger mass asymmetry.
In our previous work [16] we showed that
the mass parameter of the neck could strongly increase with mass asymmetry
starting from $\eta$=0.7 even in the hydrodynamical
treatment. In the present work we did not calculate the mass
parameters for such very asymmetric systems because our version
of the TCSM is not suitable for large $\eta$.
The mass parameter $M_{\varepsilon\varepsilon}$ increases slightly with
decreasing $\varepsilon$ or with increasing neck as shown
in Fig.~6, is almost independent of the mass number $A$ for heavy DNS
and increases with $A$. In the reactions
$^{110}$Pd+$^{110}$Pd and $^{48}$Ca+$^{172}$Hf, which
produce the same compound nucleus $^{220}$U,
the values of $M_{\varepsilon\varepsilon}$ are similar for large
$\varepsilon$, but almost two times different for small values of
$\varepsilon$.

The mass $M_{\varepsilon\varepsilon}$
becomes smaller with growing temperature as shown in Fig.~7.
The dependence of the mass parameter in neck on temperature $T_0$
is caused mainly due to the width $\Gamma_k$ of the single-particle levels
($\Gamma_k\sim T_0^2$). The value of  $M_{\varepsilon\varepsilon}$
would decrease as $T_0^{-4}$. The value of $M^{diag}_{\varepsilon\varepsilon}$
can be explained by the two-body contributions to the collective mass because
$\Gamma_k$ accounts the nucleon-nucleon collisions.
So, $M^{diag}_{\varepsilon\varepsilon}(T_0$=1 MeV)/
$M^{diag}_{\varepsilon\varepsilon}(T_0$=1.5 MeV)$\sim 5$
which is shown in Fig.~7.
One- and two-body contributions [29] are contained in
the nondiagonal and diagonal
parts of the mass parameter $M_{\varepsilon\varepsilon}$,
respectively, are obtained in the same way.
The one-body nondiagonal contribution to the
mass is relatively insensitive to
the temperature of system whereas the diagonal two-body contribution decreases
strongly with temperature. However, for the excitation energy of the
DNS formed in heavy ion collisions at low energy ($<$ 15 MeV/nucleon) the
main contribution to the mass parameter of the neck is
due to two-body components.

In Fig.~5 one can observe that the global minimum of
$M_{\lambda\lambda}$ as a function of $\beta$ in the reaction
$^{110}$Pd+$^{110}$Pd is around the ground state deformation of
$^{110}$Pd ($\beta\approx 1.2$). The mass $M_{\lambda\lambda}$
grows with decreasing elongation $\lambda$ on average as shown
in Fig.~6. The dependence of $M_{\lambda\lambda}$  on $\lambda$
has fluctuations
around an average upwards trend which are more pronounced with increasing total mass
number $A$ of the system (Fig.~6) and at smaller temperatures.
These fluctuations are related to an increasing number
of crossings between the single particle levels with decreasing $\lambda$
\cite{M1}.
The diabaticity destroys
the initial Fermi--distribution of the occupation
numbers at the touching configuration and gives rise to
fluctuating values of $\widetilde{f}_{k}$
with decreasing elongation $\lambda$ and neck parameter
$\varepsilon$ of the system.
For larger temperature, the average width $\Gamma$
increases and the function $\widetilde{f}_{k}$
becomes smoother.
Finally we note that the mass parameters
remain always much larger than those calculated with
the hydrodynamical Werner-Wheeler approximation.

\section{Summary}
Mass parameters for the relevant collective variables of
strongly deformed systems consisting of two nuclei as the DNS are
evaluated in the two-center shell model. Formulas for the masses are
derived within the linear response theory by taking the
fluctuation-dissipation theorem and
the width of single particle states
into account, and by
other methods as Fermi's golden rule and a spectral smoothing in
the mean--field cranking formula. The obtained mass
parameter for the neck degree of freedom is much larger than
the one obtained in the hydrodynamical model with the
Werner-Wheeler approximation. By applying the microscopical mass parameters
we find a relatively stable neck during
the time of reaction. In addition a large
energy threshold due to the structural forbiddenness effect \cite{11,12}
hinders the motion to proceed to smaller internuclear distances. We
observe a strong correlation between a large mass inertia and a large degree
of the structural forbiddenness for the heavy DNS as a function of
the fragmentation.  Therefore,
the DNS configuration exists a sufficiently long time and thermal
fluctuations in the mass asymmetry coordinate could develop and lead
to fusion and quasifission  which are the essential reactions
in the DNS concept. The comparison of our predicted results
in the DNS concept with many experimental data for
fusion reactions let us conclude that the DNS is  stable
against a melting in $\lambda$ and $\varepsilon$.
Our results also support the existence of long
living cluster configurations which appear in ternary
fission \cite{Ra}.

\acknowledgments
We thank Profs.
S.P.Ivanova, F.A.Ivanyuk, R.V.Jolos,
V.V.Volkov and Dr. A.B.Larionov for fruitful discussions and suggestions,
and Ms. N.G.Adamian for her help in preparing of the paper.
G.G.A. and A.D.-T. are grateful for support of the
Alexander von Humboldt-Stiftung and DAAD, respectively.
This work was supported in part by DFG and RFBR.

\begin{figure}
\centerline{\psfig{figure=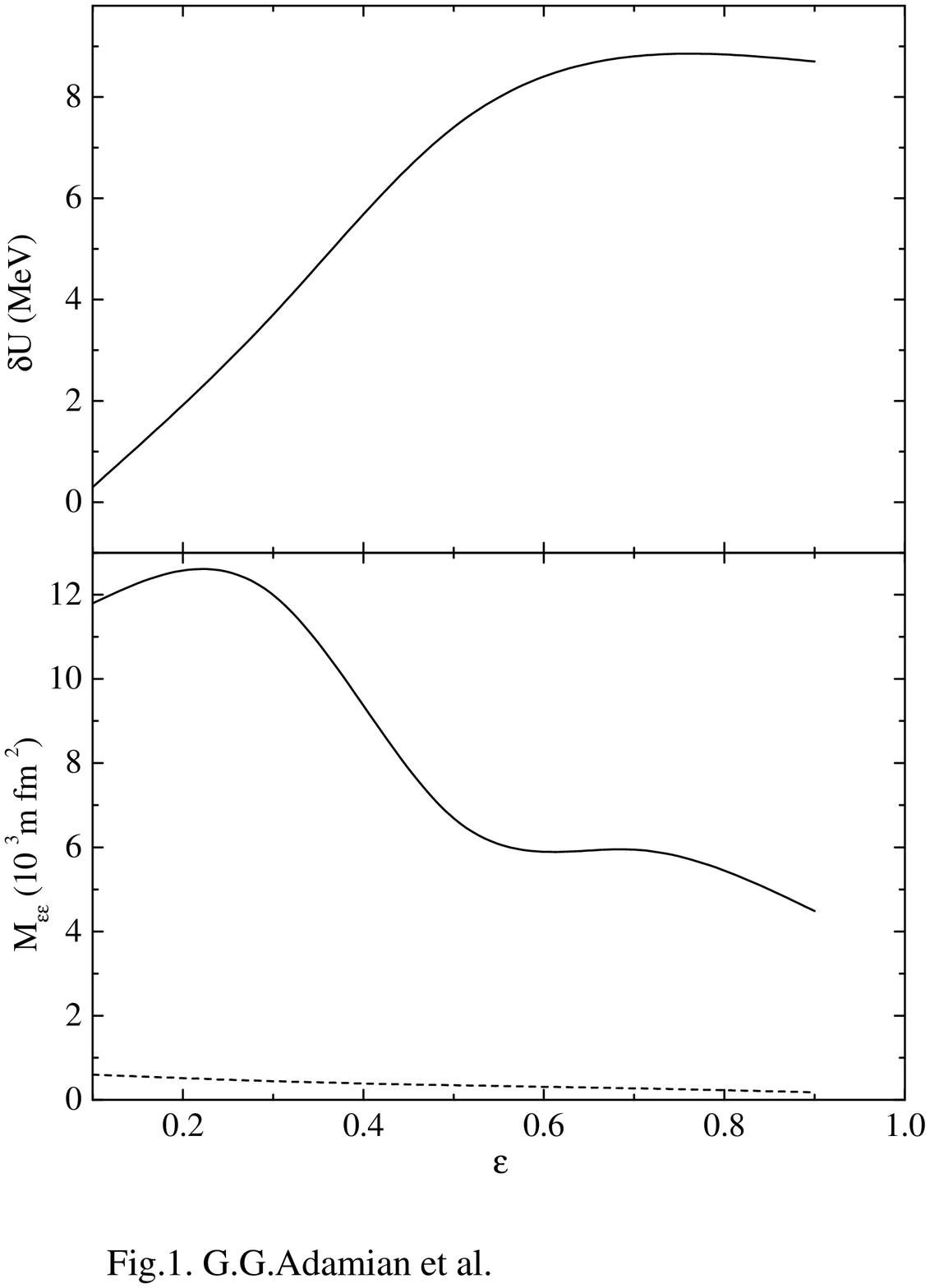,width=12.0cm}}
\caption{
Dependence of the mass parameter $M_{\varepsilon\varepsilon}$ (lower part)
and of the shell correction $\delta U$ (upper part) on $\varepsilon$
for the system $^{110}$Pd+$^{110}$Pd at $\lambda=1.6$.
In the calculation of $M_{\varepsilon\varepsilon}$
an excitation energy of 30 MeV of the DNS and adiabatic single particle states
are used. The mass parameter $M_{\varepsilon\varepsilon}^{WW}$ calculated
in the Werner-Wheeler approximation is presented by a dashed line in
the lower part.
Units: $m$ fm$^2$ with $m=$nucleon mass.
}
\label{1_fig}
\end{figure}
\newpage

\begin{figure}
\centerline{\psfig{figure=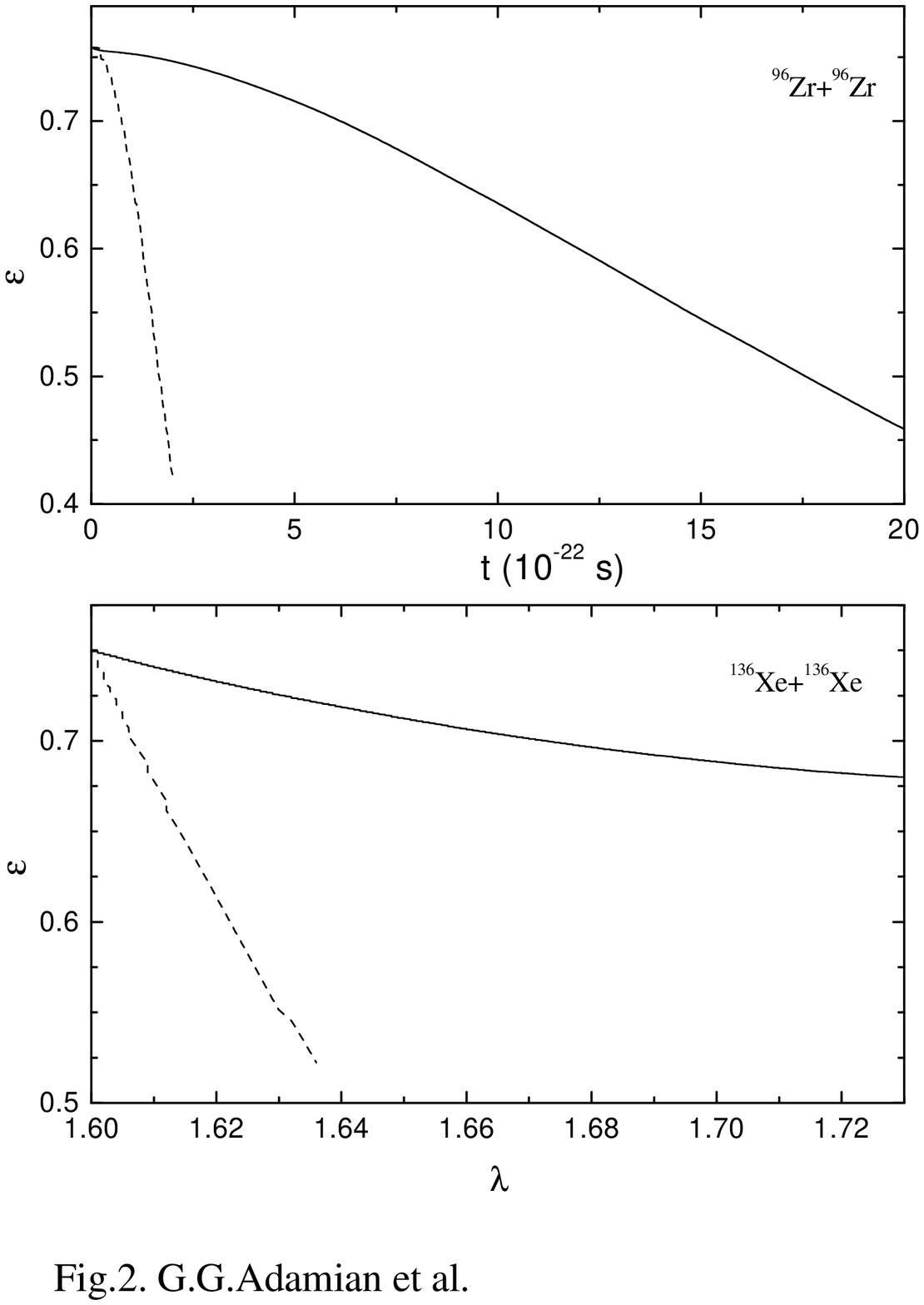,width=12.0cm}}
\caption{Upper part: Time-dependence of the neck parameter $\varepsilon$
in the system  $^{96}$Zr+$^{96}$Zr calculated with
microscopical (solid curve) and Werner--Wheeler
(dashed curve) mass parameters. Lower part:
Trajectories in the ($\lambda$,$\varepsilon$)-plane
calculated for the system  $^{136}$Xe+$^{136}$Xe
with microscopical mass parameters
(solid curve) and with Werner--Wheeler mass parameters
(dashed curve). The end points of the solid and dashed curves in
the drawing are at time $t=2\times10^{-21}$s and
$t=2\times10^{-22}$s, respectively.}
\label{2_fig}
\end{figure}
\newpage

\begin{figure}
\centerline{\psfig{figure=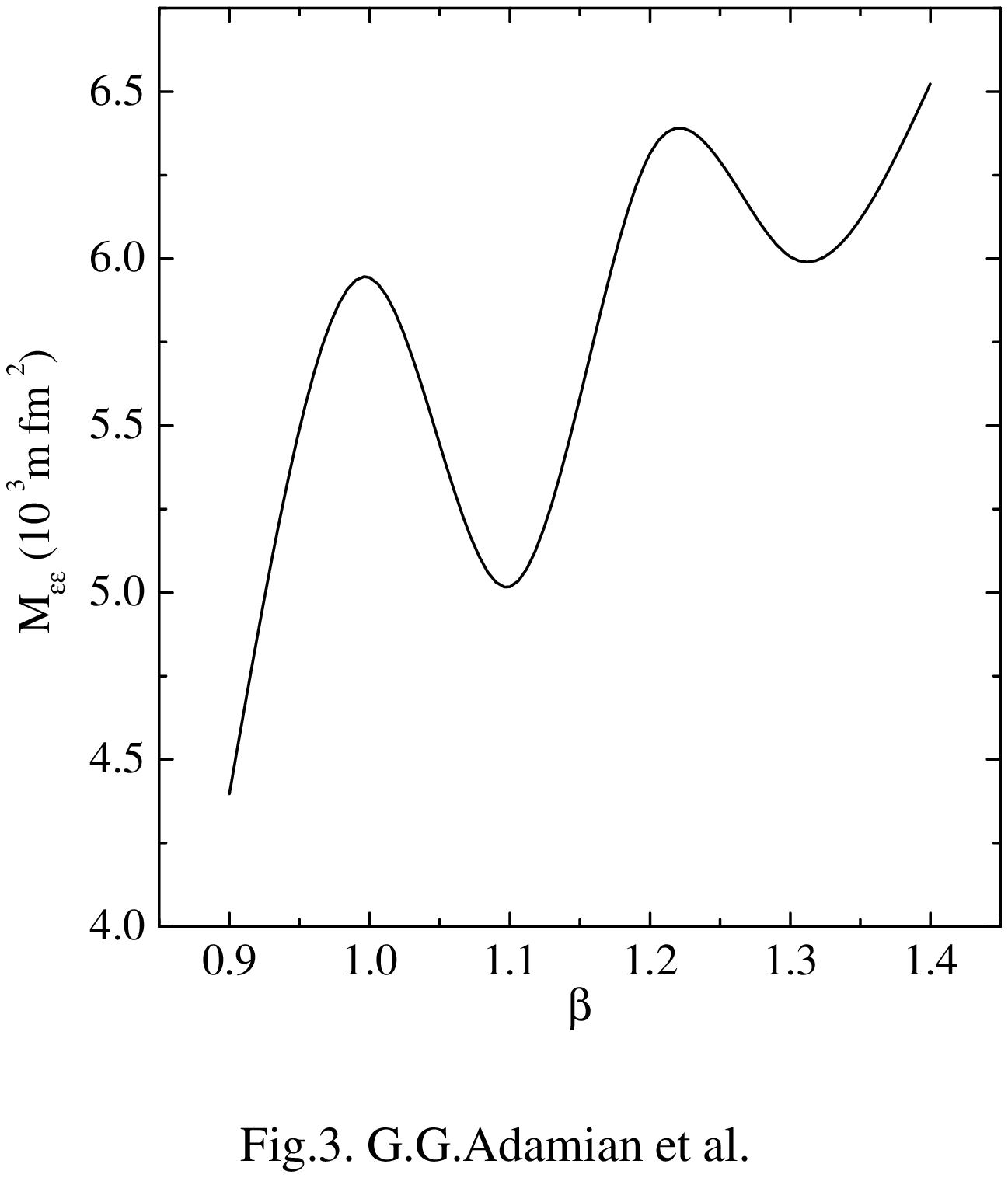,width=12.0cm}}
\caption{ Mass parameter  $M_{\varepsilon\varepsilon}$
as a function of
deformation $\beta$ calculated for the system
$^{110}$Pb+$^{110}$Pb at the touching configuration
with excitation energy 30 MeV and adiabatic single particle
states.
Units: $m$ fm$^2$ with $m=$nucleon mass.
}
\label{3_fig}
\end{figure}
\newpage

\begin{figure}
\centerline{\psfig{figure=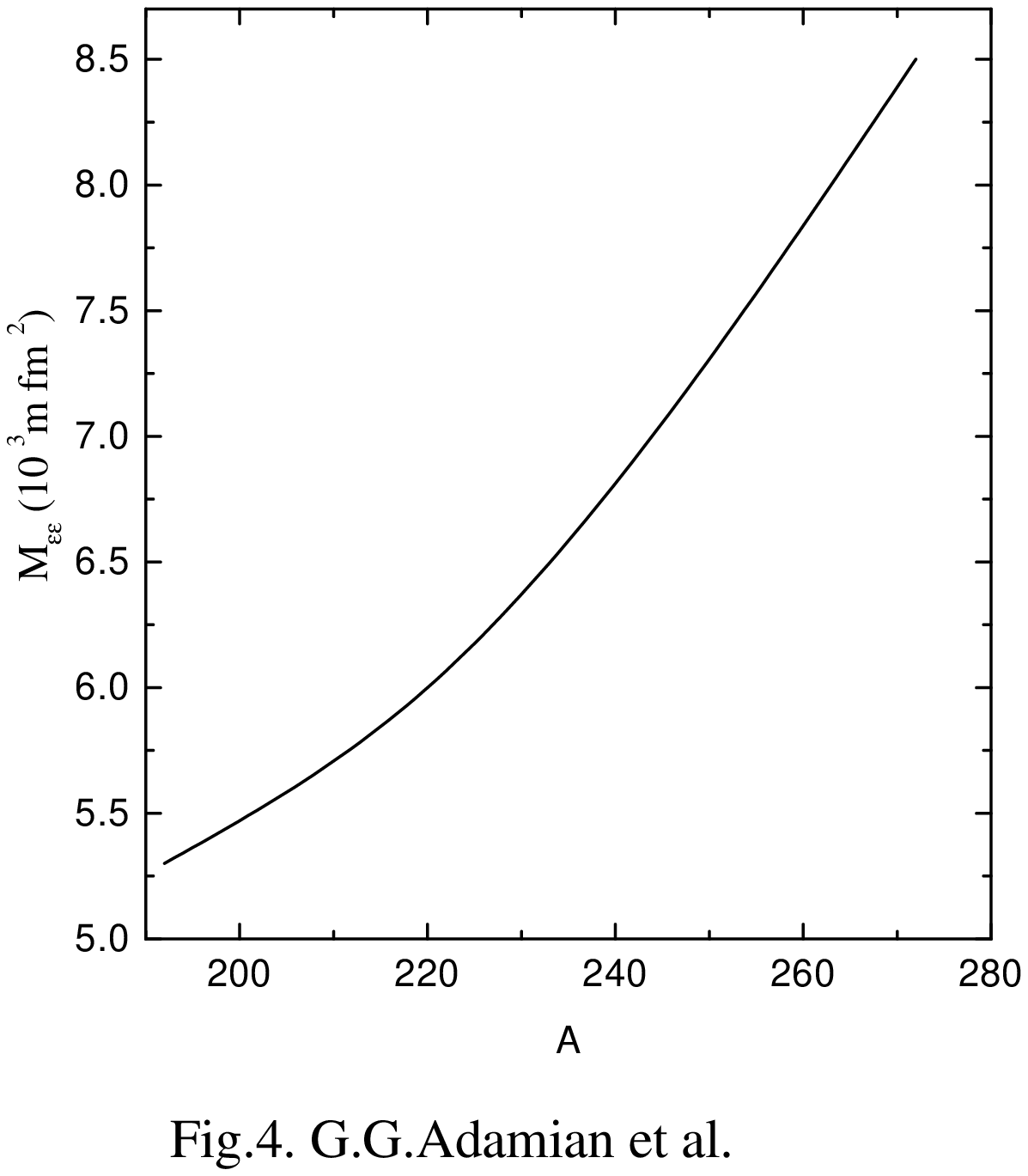,width=12.0cm}}
\caption{ Mass parameter $M_{\varepsilon\varepsilon}$
as a function of the mass number $A$ of symmetric systems ($\eta=0$)
calculated with adiabatic single particle states
 for $\beta_i=1$, $\lambda=1.6$,
and $\varepsilon=0.75$.
Units: $m$ fm$^2$ with $m=$nucleon mass.
}
\label{4_fig}
\end{figure}
\newpage

\begin{figure}
\centerline{\psfig{figure=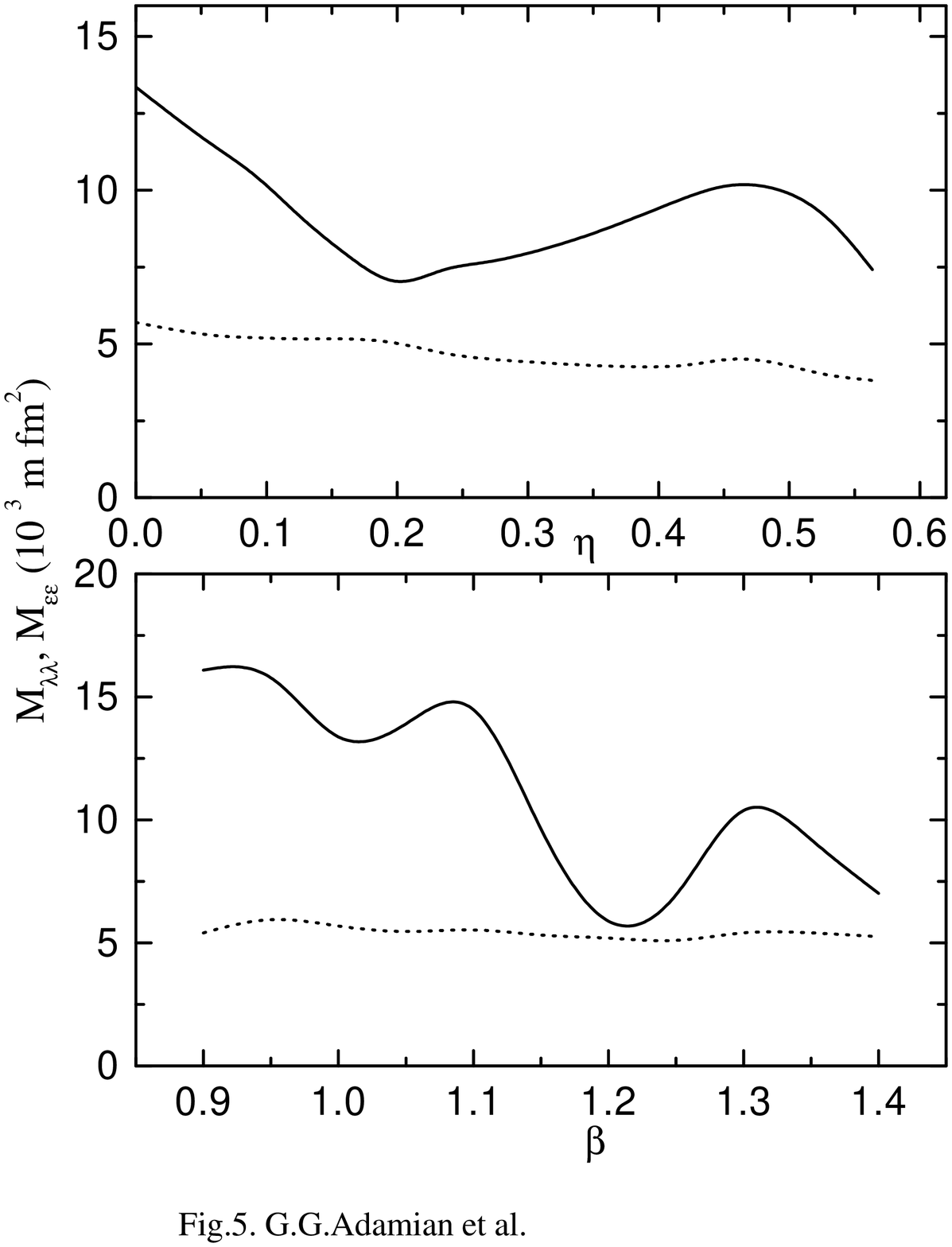,width=12.0cm}}
\caption{
Upper part: Mass parameters   $M_{\lambda\lambda}$ (solid line)
and  $M_{\varepsilon\varepsilon}$ (dotted line) as functions of mass asymmetry
$\eta$ at the touching configuration for DNS which leads to the same
compound nucleus $^{220}$U. The value of the neck coordinate is
$\varepsilon$=0.75 and excitation energy of the DNS is 30 MeV.
The calculation is done with spherical nuclei.
Lower part: The same as in the upper part, but as a function of
deformation $\beta$ for the reaction
$^{110}$Pd+$^{110}$Pd ($\eta=0$). Diabatic single particle states are used.
Units: $m$ fm$^2$ with $m=$nucleon mass.
}
\label{5_fig}
\end{figure}
\newpage

\begin{figure}
\centerline{\psfig{figure=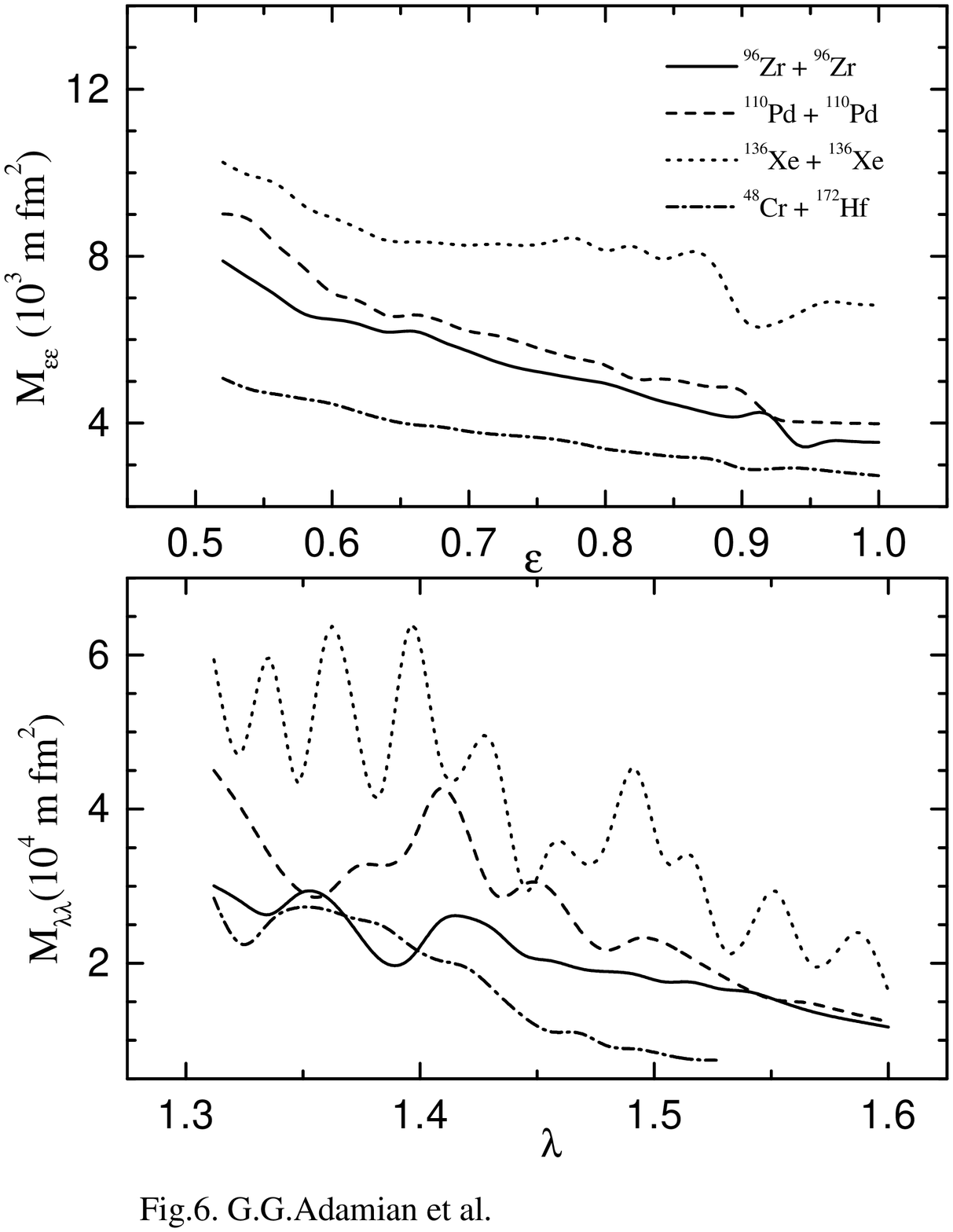,width=12.0cm}}
\caption{
Upper part: Mass parameter  $M_{\varepsilon\varepsilon}$
as a function of the neck coordinate
$\varepsilon$ at the touching configuration
in the reactions:
$^{96}$Zr+$^{96}$Zr (solid line), $^{110}$Pd+$^{110}$Pd
(dashed line),
$^{136}$Xe+$^{136}$Xe (dotted line)
and
$^{48}$Ca+$^{172}$Hf (dashed--dotted line).
The  excitation energy of the DNS
in these reactions is 30 MeV.
Lower part: The same as in the upper part, but
for $M_{\lambda\lambda}$
as a function of the
elongation $\lambda$. Diabatic single particle states are used.
Units: $m$ fm$^2$ with $m=$nucleon mass.
}
\label{6_fig}
\end{figure}
\newpage

\begin{figure}
\centerline{\psfig{figure=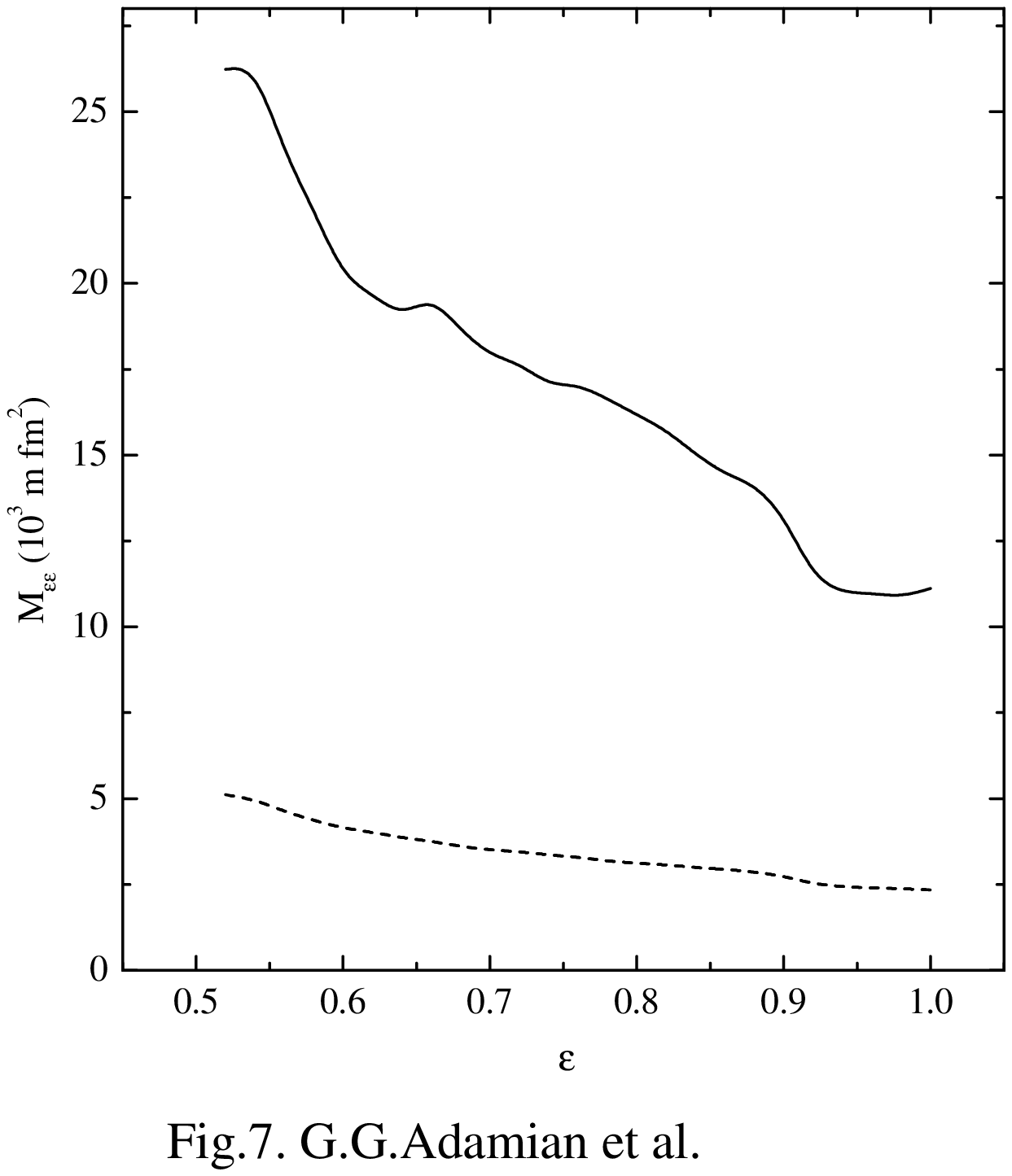,width=12.0cm}}
\caption{ Mass parameter $M_{\varepsilon\varepsilon}$
as a function of $\varepsilon$ at the touching configuration
in the $^{110}$Pd+$^{110}$Pd reaction for
temperatures $T_0=1$ MeV (solid line) and
$1.5$ MeV (dashed line). Diabatic single particle states are used.
Units: $m$ fm$^2$ with $m=$nucleon mass.
}
\label{7_fig}
\end{figure}

\end{document}